\documentclass[submission,copyright,creativecommons]{eptcs}
\providecommand{\event}{PxTP 2021} 
\usepackage{breakurl}             
\usepackage{underscore}           

\newcommand\vtrue{{\sf true}}
\newcommand\vfalse{{\sf false}}

\title{Proof Generation in CDSAT}
\author{Maria Paola Bonacina
\institute{Dipartimento di Informatica,
Universit\`a degli Studi di Verona,
Verona, Italy}
\email{mariapaola.bonacina@univr.it}}
\def\titlerunning{Proof Generation in CDSAT}
\def\authorrunning{M. P. Bonacina}
\begin{document}
\maketitle

Proofs of unsatisfiability of a negated conjecture, or, equivalently, proofs of validity
of the original conjecture, are an essential output of automated reasoning methods.
The transformation, exchange, and standardization of proofs
is a key factor for the interoperability of different automated reasoning systems.
In theorem proving {\em proof reconstruction} is the task of extracting a proof from the
final state of a derivation after generating the empty clause.
While for several theorem proving methods and theorem provers it is a standard task,
it is never trivial. For example,
in parallel theorem proving with distributed search
(see \cite{MPB:PCRHandbook:2018:PTP} for a recent survey),
multiple parallel processes perform inferences and search for a proof.
A parallel theorem proving method has {\em distributed proof reconstruction},
if the process that generates the empty clause can reconstruct the proof
from the final state of its database,
even if all processes contributed to the proof \cite{MPB:JSC:1996:mcd}.

In propositional satisfiability (SAT) solving,
the {\em conflict-driven clause learning} (CDCL) procedure generates proofs by {\em resolution},
because it uses resolution to explain conflicts \cite{ZhangMalik:DATE2003,ShankarSurvey:2009}.
SAT solvers apply pre-processing steps
and simplification techniques that also need to be accounted for in proofs.
Furthermore, proofs generated by SAT solvers are so huge that
their definition, generation, and manipulation, involving various proof formats,
is an important research topic (e.g., \cite{CHHKS:CADE2017:RAT}).

Satisfiability modulo theories (SMT) solving
represents a middle ground between first-order theorem proving and SAT solving.
Initially, {\em model generation} was emphasized over proof generation in SMT,
because the focus was on fragments of first-order theories where satisfiability is decidable,
in contrast with first-order logic where satisfiability is not even semidecidable.
Over time, SMT solvers have been applied more and more to unsatisfiable inputs,
including inputs with quantifiers that may fall outside decidable fragments.
SMT solvers have become more similar to theorem provers,
and proof generation is crucial also in SMT.
Since most SMT solvers are built on top of the CDCL procedure, their proofs are proofs by resolution
(with the same caveat as above)
with proofs of theory lemmas plugged in as leaves or {\em black-box sub-proofs}
\cite{FMMNT:TACAS:2006,BdM:IWIL2008,MPB-MKJ:JAR:2015:interpolationGround,KBTRH:FMCAD2016,BBFF:JAR2020}.

CDSAT ({\em Conflict-Driven SATisfiability}) is a paradigm for SMT
that innovates SMT solving in several ways
\cite{MPB-SGL-NS:CADE2017:CDSAT,MPB-SGL-NS:CPP2018:CDSAText,MPB-SGL-NS:JAR:2019:CDSATtransition,MPB-SGL-NS:JAR:CDSAText}.
To begin with, CDSAT solves {\em SATisfiability problems Modulo theories and Assignments} (SMA),
which means that the input problem may contain assignments to first-order terms (e.g., $x\gets 3$).
The solver has to determine whether there exists a model that satisfies the input
formula and also {\em endorses} the input first-order assignments.
A model {\em endorses} an assignment if it interprets identically left hand side and right hand side of
the assignment.
For uniformity, CDSAT views also formul\ae\ as assignments to Boolean terms
(e.g., $(\neg A \vee B)\gets \vtrue$), and seeks a model that endorses all input assignments.
There is a subtle technical difference between, say, $x\gets 3$ and
$(x\simeq 3)\gets \vtrue$, since in the latter $3$ is a constant symbol of the input language,
whereas in the former $3$ is a {\em value}, whose denotation requires a {\em theory extension}.
The generalization of SMT to SMA is relevant to approaching {\em optimization} problems
by solving iteratively SMA problems, where input first-order assignments are used to exclude
sub-optimal solutions and induce a convergence towards an optimal one \cite{dMP:ADDCT2013}.

As the name says, CDSAT is a {\em conflict-driven} method.
In general, a procedure is {\em conflict-driven} if it proposes a candidate model represented by a series of
assignments, and performs non-trivial inferences only to explain a conflict between the current
candidate model and the formul\ae\ to be satisfied.
Since in CDSAT also formul\ae\ are assignments, the separation between candidate model
and formul\ae\ disappears.
The state of the computation is simply a sequence of assignments $\Gamma$, called a {\em trail},
which also contains the input assignments.
A {\em conflict} is a subset of $\Gamma$ that is unsatisfiable.

CDSAT is designed since the start for reasoning in a {\em union of theories},
with propositional logic as one of the theories.
CDSAT lifts the conflict-driven style of CDCL from propositional logic
to {\em conflict-driven reasoning in a union of theories}; and it
reduces to CDCL if propositional logic is the sole theory.
Prior to CDSAT, MCSAT ({\em Model-Constructing SATisfiability})
\cite{dMJ:VMCAI2013,JBdM:FMCAD2013,ZWR:SAT2016,J:VMCAI2017,BGLMB:SMT2018,SGL-DJ-BD:IJCAR2020}
showed how to integrate CDCL with a conflict-driven theory satisfiability procedure
(e.g., \cite{JdM:IJCAR2012,JdM:JAR:2013,BKKM:FroCoS2019}
and see \cite{MPB:AFM-NFM2017:conflictres} for a survey with more references).
CDSAT generalizes MCSAT to generic unions of {\em disjoint} theories,
meaning that their signatures do not share symbols other than equality on shared sorts.
CDSAT resembles MCSAT, if there are only propositional logic
and another theory with a conflict-driven satisfiability procedure.

For an input problem to be satisfiable in a union of theories,
the theories need to agree on which shared terms are equal and on the cardinalities of shared sorts.
Beginning with the pioneering work of Nelson and Oppen
\cite{NelsonOppen:TOPLAS:1979,Nelson:1983},
most approaches to reasoning in a union of theories are defined as
{\em combination schemes} that combine theory satisfiability procedures
(see \cite{MPB-PF-CR-CT:LNAI:2019:BeyondES} for a survey with more references).
These schemes {\em separate} the original problem into sub-problems, one per theory in the union.
The completeness of the combination scheme rests on a {\em combination lemma}
that states which conditions the theories need to satisfy in order to agree
on the cardinalities of shared sorts.
The satisfiability of the original input in the union of theories is reduced to the satisfiability
of every sub-problem in the respective theory,
where every sub-problem is conjoined with an {\em arrangement}.
An {\em arrangement} is a conjunction of equalities and disequalities between shared variables,
or shared constants, depending on whether free variables or constants are used to represent shared terms.
In a non-deterministic description the arrangement can be guessed.
In practice, it is computed by the theory satisfiability procedures.
The computation of the arrangement is the only activity where the theory satisfiability procedures
cooperate, typically by exchanging equalities between shared variables.

In contrast with this traditional setting,
CDSAT is defined as a {\em transition system} that orchestrates theory-specific {\em inference systems},
called {\em theory modules}. An inference system is a set of inference rules,
and a theory module is an abstraction of a satisfiability procedure.
Every module has its view of the trail, called {\em theory view},
which contains whatever the module can understand.
A theory module can {\em expand} the trail with an assignment that is a {\em decision},
encapsulated in the {\tt decide} transition rule of CDSAT,
or the result of a {\em theory inference},
encapsulated in the {\tt deduce} transition rule of CDSAT.
Theory inferences are used for propagations, and conflict detection
and explanation in the respective theory.
The latter applies until the theory conflict surfaces on the trail as a Boolean conflict
(e.g., $L\gets\vtrue$ and $\neg L\gets\vtrue$, or, equivalently, $L\gets\vtrue$ and $L\gets\vfalse$).
Then the conflict-solving transition rules of CDSAT come into play.
Since the Boolean conflict may descend from first-order assignments,
the conflict-solving transition rules of CDSAT are designed to handle both Boolean and first-order
assignments. It does not matter whether a theory satisfiability procedure is conflict-driven,
because CDSAT is conflict-driven for all theories.
At the very least, a theory satisfiability procedure can be abstracted into a {\em black-box theory module},
with an inference rule that detects unsatisfiability by invoking the procedure.

The completeness of CDSAT rests mainly on properties of the theory modules.
Every theory module is required to be {\em complete}, meaning that
it can expand its view of the trail if it is not satisfied by a model of its theory.
One of the theories in the union needs to be the {\em leading theory}.
A leading theory is aware of all the sorts in the union of theories,
and its theory module is aware of all the constraints that the theories may have
on the cardinalities of shared sorts.
While for the leading theory module it suffices to be complete,
any other module needs to be {\em leading-theory-complete},
meaning that it can expand its view of the trail if it is not satisfied by a model of its theory
that concurs with a model of the leading theory on cardinalities of shared sorts
and equality of shared terms.

This description shows that while the traditional combination schemes combine decision procedures
as {\em black-boxes}, CDSAT provides a tighter form of integration at the inference level.
This has consequences on {\em proof generation}.
Since the conflict-driven reasoning happens directly in the union of the theories
and not only in propositional logic, resolution does not have a dominant role.
CDSAT proofs can be rendered as resolution proofs,
but this is not a necessary choice.
Since the theory satisfiability procedures are not combined as black-boxes,
theory sub-proofs are not necessarily black-boxes either.
Since CDSAT solves SMA problems, also first-order assignments may appear in proofs.
The theory inferences may introduce {\em new} (i.e., non-input) terms,
in order to explain conflicts.
Thus, such new terms may appear in proofs.

The powerful abstractions that characterize CDSAT leads to proof generation approaches
also based on abstraction.
The CDSAT transition system can be made {\em proof-carrying},
by equipping the transition rules with the capability to generate {\em abstract proof terms}.
During proof reconstruction these proof terms can be translated into different proof formats,
including resolution proofs.
The resulting proofs can be dispatched to proof checkers or proof assistants,
or otherwise manipulated and integrated.
Alternatively, CDSAT can adopt the LCF style for proofs,
which avoids building proof objects in memory altogether.
In LCF style, the prover or solver (e.g., a CDSAT based solver)
is built on top of a {\em trusted kernel} of primitive operations.
When the reasoner detects unsatisfiability, the refutation is correct by construction,
because otherwise a type error would arise.

\paragraph{Acknowledgements}
The author thanks St\'ephane Graham-Lengrand for their discussions.

\nocite{*}
\bibliographystyle{eptcs}
\bibliography{myBib}
\end{document}